\documentclass{article}
\pdfoutput=1
\usepackage{jheppub}

\usepackage{amssymb,amsmath,amsfonts,amsbsy,epsfig,wrapfig,caption,cancel,graphicx, pstricks, slashed, mathtools,xcolor}

\usepackage{babel}
\usepackage[latin9]{inputenc}
\usepackage{float}
\usepackage{hyperref}
\usepackage[mathscr]{eucal}
\usepackage{esint}
\usepackage{amsbsy}
\usepackage{color}
\usepackage{cleveref}
\usepackage{breakurl}
\usepackage{ulem}
\usepackage{bbm}
\usepackage{mathtools}
\usepackage{braket}
\usepackage{comment}
\usepackage[super]{nth}
\usepackage{hyphenat}

\newcommand{\eq}[2]{\begin{equation}\label{#1}#2 \end{equation}}

\usepackage{xcolor}

\newcommand{\ep}{\epsilon}
\newcommand{\la}{\lambda}
\newcommand{\p}{\varphi}
\newcommand{\g}{\gamma}

\newcommand{\calO}{{\cal O}}

\def\BSM{\left(\begin{smallmatrix}}
\def\ESM{\end{smallmatrix}\right)}
\def\BM{\left(\begin{matrix}}
\def\EM{\end{matrix}\right)}

\def\t{$_2$}
\def\vf{\varphi}
\def\d{\partial}

\def\bea{\begin{eqnarray}}
\def\eea{\end{eqnarray}}
\def\be{\begin{equation}}
\def\ee{\end{equation}}
\def\ba{\begin{array}}
	\def\ea{\end{array}}

\def\hhh{{\cal H}}
\def\tz{\tau_0}
\def\tvv{\tau_1}
\def\tvs{\tau_2}

\definecolor{Dgreen}{rgb}{0,0.7,0.0}

\begin{document}

\title{\Large{Backreaction of Schwinger pair creation in massive QED$_2$}}

\author[a]{Gregory Gold,}
\affiliation[a]{Niels Bohr International Academy, Niels Bohr Institute,\\ Blegdamsvej 17, Copenhagen, DK 2100, Denmark\\}

\author[b]{David A. McGady,}
\affiliation[b]{Nordita, KTH Royal Institute of Technology and Stockholm University, \\
	Roslagstullsbacken 23, SE-106 91 Stockholm, Sweden\\}

\author[a, c]{Subodh P. Patil}
\affiliation[c]{Instituut-Lorentz for Theoretical Physics, Leiden University,\\ 2333 CA Leiden, The Netherlands\\}

\author[d]{and Valeri Vardanyan}
\affiliation[d]{Kavli Institute for the Physics and Mathematics of the Universe (WPI), UTIAS,\\ The University of Tokyo, Chiba 277-8583, Japan\\}

\emailAdd{greg.gold1@gmail.com, dmcgady@alumni.princeton.edu, patil@lorentz.leidenuniv.nl, valeri.vardanyan@ipmu.jp}

\date{\today}

\abstract{Particle-antiparticle pairs can be produced by background electric fields via the Schwinger mechanism provided they are unconfined. If, as in QED in (3+1)-$d$ these particles are massive, the particle production rate is exponentially suppressed below a threshold field strength. Above this threshold, the energy for pair creation must come from the electric field itself which ought to eventually relax to the threshold strength. Calculating this relaxation in a self-consistent manner, however, is difficult. Chu and Vachaspati addressed this problem in the context of capacitor discharge in massless QED$_2$~\cite{Chu:2010xc} by utilizing bosonization in two-dimensions. When the bare fermions are massless, the dual bosonized theory is free and capacitor discharge can be analyzed exactly~\cite{Chu:2010xc}, however, special care is required in its interpretation given that the theory exhibits confinement. In this paper we reinterpret the findings of \cite{Chu:2010xc}, where the capacitors Schwinger-discharge via electrically neutral dipolar meson-production, and generalize this to the case where the fermions have bare masses. Crucially, we note that when the initial charge of the capacitor is large compared to the charge of the fermions, $Q \gg e$, the classical equation of motion for the bosonized model accurately characterizes the dynamics of discharge. For massless QED$_2$, we find that the discharge is suppressed below a critical plate separation that is commensurate with the length scale associated with the meson dipole moment. For massive QED$_2$, we find in addition, a mass threshold familiar from (3+1)-$d$, and show the electric field relaxes to a final steady state with a magnitude proportional to the initial charge. We discuss the wider implications of our findings and identify challenges in extending this treatment to higher dimensions.}

\maketitle

\section{Preliminary remarks}
The most remarkable thing about the vacuum is that it conducts whenever the applied potential difference across a given region exceeds a critical value -- an effect first identified in \cite{Sauter:1931zz, Heisenberg:1935qt} that now bears the name of Schwinger \cite{Schwinger:1951nm}. Schwinger gave a complete characterization of how positrons and electrons can be pair produced once the applied electric field in a given region exceeds the critical strength
\eq{SL}{E_c = \frac{m_e^2 c^3}{e \hbar} \approx 10^{18} \rm{V/m}.}
This is also the field strength above which quantum effects render electromagnetism to be non-linear. This raises the immediate question, where does the energy required to produce these particle-antiparticle pairs come from? Clearly, it must be drawn from the electric field itself, which must subsequently decay until it reaches the threshold strength Eq.~\eqref{SL} when pair production becomes exponentially suppressed again. However, quantifying how this occurs is a challenging problem to say the least. 

Schwinger's original calculation relied on background field quantization, wherein the electric field is taken to be an external field and the 1PI effective action is evaluated in the proper time formalism by integrating out the fermion fields. On a background with a constant electric field, this results in an imaginary contribution to the effective action (see for instance \cite{Dunne:1998ni})
\eq{}{{\rm Im \,S} =  \frac{(e E)^2}{8\pi^3}V \sum_{n = 1} \frac{1}{n^2} e^{-\frac{m^2 \pi n}{e E}},}
where we work in natural units, and where $V = L^3 T$ is the volume of spacetime. 

In this context, the imaginary part of the effective action corresponds to a breakdown of the vacuum state which contains neither electrons nor positrons. Put differently, the imaginary component of the one-loop effective action corresponds to a breakdown of the fermionic vacuum state via electron-positron pair creation. Nikishov~\cite{Nikishov:1970br} and others, e.g.~\cite{Cohen:2008wz}, have emphasized that the actual rate of pair creation per unit volume of spacetime is just (twice) the first term in this infinite sum:
\eq{pp}{ \Gamma = \frac{dN}{dt dV} = \frac{(e E)^2}{4\pi^3} e^{-\frac{\pi m^2}{e E}}~.} 
These computations are done in the strict semiclassical limit using background field quantization, where the charged particles produced by the externally applied electromagnetic fields have negligible effect on the external fields themselves. In other words, these computations explicitly ignore the dynamics of backreaction.

However, it is very important to understand how this backreaction process works in detail, both in the particular context of Schwinger-dissipation and in the much more general context of quantum back-reactions induced by classical field configurations dissipating the initial classical solutions themselves \cite{Fradkin:1991zq}. In this context, the Schwinger mechanism, where an initial external electric field causes back-reaction against itself, is a comparatively simpler process than analogous back-reaction processes in (quantum) gravity such as black-hole evaporation. Further, as Schwinger pair creation may actually be observed in laboratory settings such as single-layer graphene~\cite{Allor:2007ei}, it is important to have a realistic, concrete, theoretical understanding for how such systems will evolve in time. 

There are a number of ways one could imagine calculating the backreaction on the electric field of the produced particles. The brute force method would be to interpret Eq.~\eqref{pp} as the energy extracted per unit spacetime volume from a spatially uniform electric field and make an ansatz for a new time dependent but spatially uniform electric field, which would also source a time dependent magnetic field, the sum of whose energy densities are required to dissipate according to Eq.~\eqref{pp}. The vector potential one calculates from this ansatz would be the new background which one would background field quantize around to calculate the imaginary part of the effective action. One then repeats this process iteratively in the hope that this procedure converges on a self consistent quantum corrected background -- a process that can quickly become cumbersome to the point of being impractical.

Evidently, the problem of completely characterizing the dissipation of the background electric field from particles produced via the Schwinger effect lies within the ambit of non-equilibrium techniques, however the precise manner in which one should proceed is unclear.\footnote{For instance, 2PI methods and their possible extensions (see \cite{Berges:2004yj} for an excellent review) are problematic from the perspective of maintaining gauge invariance in QED \cite{Mottola:2003vx}.} In spite of this, some notable attempts have been made using a variety of approximations. In \cite{Akhmedov:2009vs}, the authors made a truncation of the recursive procedure discussed above in the context of the Euler-Heisenberg effective action for scalar QED. Notable also is the mean field approach in the context of scalar QED of \cite{Habib:1995ee}. Earlier, the authors of \cite{Cooper:1989kf,Kluger:1991ib,Kluger:1992gb} adopted a semi-classical approach that replaced currents with their expectation values and attempted a numerical study of the backreaction problem; here particle production and back-reaction is studied with the use of Boltzmann-Vlasov equations (see also \cite{Kluger:1998bm}). The effects of time varying external fields in different spacetime dimensions was studied in \cite{Gavrilov:1996pz}, and the study of secular growth of loop corrections in the proper time formalism was studied and carefully interpreted in \cite{Akhmedov:2014hfa, Akhmedov:2014doa}. However a fully self-consistent background at the quantum level remains as of yet elusive. In a remarkable insight, Chu and Vachaspati \cite{Chu:2010xc} realized the latter could in fact be addressed by considering the problem of vacuum capacitor discharge, albeit in the context of 1+1-dimensions in a theory with massless fermions -- massless QED\t -- by considering its bosonized version, an approach we first review before reinterpreting and extending to the case of massive bare fermions. 

The first obstacle one must address in the interpretation of~\cite{Chu:2010xc} is that in 1+1-dimensions, both massless and massive QED\t ~exhibit confinement, and so any interpretation of the electric field discharge in terms of currents as put forward in \cite{Chu:2010xc} must be treated with care. We find by considering the duality between the massless Schwinger model \cite{Schwinger:1962tp} and the free (gapped) dual scalar field theory, the electric field discharge found in \cite{Chu:2010xc} should be interpreted in terms of meson production, whereby bound particle-antiparticle pairs are produced whenever the plate separation surpasses the threshold length scale defined by the electric dipole moment of the produced mesons. Capacitor discharge then proceeds via dipole screening by the induced meson cloud. In this way, we can qualify one of the caveats usually applied in applying lessons learned from studying Schwinger pair production towards Hawking evaporation or the semi-classical stability of de Sitter space \cite{dS1, Anderson:2017hts} -- namely that such interpretations are limited by the fact that there are no negative gravitational charges (i.e. masses). Here, we show in the context of a toy model that \textit{electric field discharge can proceed even with the production of neutral particles.\footnote{However, this qualification is immediately tempered by the fact that gravitational dipoles do not exist either.}}

The structure of this paper is as follows. In section~\ref{secMassless}, we first describe the mapping between massless QED\t~and the integrable bosonized theory and then revisit the analysis of~\cite{Chu:2010xc} applied to the problem of electric field discharge within a capacitor. Importantly, we identify additional characteristic timescales to that noted in~\cite{Chu:2010xc}. In section~\ref{secMeson}, we physically interpret this discharge. Our analysis tracks the content of~\cite{Coleman:1975pw} and~\cite{Coleman:1976uz} closely by focusing on the properties of the physical excitations of the system, i.e. the \textit{massive} and electrically neutral $\vf$-mesons which carry electric dipole moments, identifying a threshold plate separation below which discharge does not occur. We further identify additional timescales associated with dynamical capacitor discharge in massless QED\t. Then, in section \ref{secMassive1}, we explicitly show how bare fermion masses change the relevant equations of motion. Crucially, adding a fermion-mass induces a non-zero cosine-potential within the dual bosonic theory. In section~\ref{secMassive2}, we study the equation of motion for the bosonic configuration dual to capacitor-discharge in massive QED\t. When the charge of the capacitor plate is large compared to the charge of the massive fermion in QED\t, i.e. when $Q \gg e$, occupation numbers are large and we can self-consistently integrate the classical non-linear equations of motion on the bosonic side, recovering a mass threshold at the critical field $E_c = \frac{ e^{\gamma_E}m_\psi}{2\sqrt{\pi}}$, where $\gamma_E$ is the Euler gamma constant and $m_\psi$ is the bare fermion mass in the notation of section \ref{secMassive1}. As we discuss in section \ref{secMassive2}, the bare mass $m_\psi$ is not to be taken as the mass of any physical state in the massive Schwinger model, rather it is to be interpreted as a parameter that determines the spectrum of states in the theory. We conclude by comparing the final discharge profile in massive and massless QED\t, highlighting the role of the two independent thresholds as we dial the plate separation and initial charge on the capacitors. 

\section{Integrability of capacitor discharge in massless QED\t}\label{secMassless}

In this section, we review the treatment of~\cite{Chu:2010xc} which analytically computed capacitor discharge in massless QED\t. Our purpose is twofold. First, so that we can revisit the physical interpretation of Ref.~\cite{Chu:2010xc}'s results for capacitor discharge, in light of the mass-gap due to electron confinement within the Schwinger model~\cite{Schwinger:1962tp,Coleman:1975pw}. Second, to lay the groundwork for our study of capacitor discharge in 1+1-dimensions with massive charged fermions in sections~\ref{secMassive1} and~\ref{secMassive2}. 

To begin we note that the insight of \cite{Chu:2010xc} was to point out that the action for massless QED in 1+1-dimensions,
\begin{align}
S_{\psi,A} := \int d^2x \bigg( \bar\psi ( i \slashed D - e \slashed A) \psi - \frac{1}{4} F_{\mu \nu} F^{\mu \nu} \bigg) ~~,
\end{align}
and the action tying a putatively massless scalar to the same gauge-field and its corresponding field strength, $A_{\mu}$ and $F_{\mu \nu} := \d_\mu A_\nu -  \d_\nu A_\mu$, 
\begin{align}
S_{\vf,A} := \int d^2x \bigg( (\d^{\mu} \p)(\d_{\mu} \p) + g \vf \big(\ep^{\mu \nu} F_{\mu \nu}\big) - \frac{1}{4} F_{\mu \nu} F^{\mu \nu} \bigg)~~,
\label{eqSphiA}
\end{align}
where $g := \frac{e}{\sqrt{\pi}}$, describe the same theory at the quantum mechanical level via the following mapping between normal-ordered (:$AB$:) operators in $S_{\vf,A} \leftrightarrow S_{\psi,A}$:
\begin{align}
: \bar \psi i \g^{\mu} \psi : \leftrightarrow \ep^{\mu \nu} \d_{\nu} \p~ \implies S_{\psi,A} \leftrightarrow S_{\vf,A}~~.
\end{align}
This operator correspondence, an important ingredient in the duality between the massive-Thirring model and the sine-Gordon model~\cite{Coleman:1974bu}, is the principle ingredient that allowed Ref.~\cite{Chu:2010xc} to analytically solve for the dynamics of capacitor discharge in massless QED in 1+1-dimensions.

Fundamentally, this operator map allows the trilinear coupling in ${\rm QED}_2$ to be converted into a quadratic coupling in the dual theory. Up to topological boundary terms, we have :$\bar \psi \slashed A \psi$:$\leftrightarrow\!\!\vf \ep^{\mu \nu} F_{\mu \nu}$. This, in turn, converts equations of motion that are quadratic in $\psi$ into linear equations of motion for $\p$. We then consider the $\p$-field profile that corresponds to an initial state corresponding to a parallel-plate capacitor in QED. Integrating the equations of motion for $\p$~\cite{Chu:2010xc} gives closed form expressions for the charge density, i.e. $:\bar \psi \g^{\mu} \psi:$, and thus yields an exact description of capacitor discharge. 

However, one must be mindful of a number of conceptual points in the mapping between the actions $S_{\psi,A}$ and $S_{\vf,A}$, relevant to both massless QED\t~\cite{Chu:2010xc} and our extension to massive QED\t. Primarily, as Schwinger showed in 1962~\cite{Schwinger:1962tp}, QED in 1+1-dimensions with \textit{massless} fermions is known to have a gapped spectrum. Further, there is no ``photon'', that is, electromagnetic fields are non-dynamical in 1+1-dimensions. Thus any putatively dynamical gauge-field $A^{\mu}$ may be integrated out of any on-shell description of the \textit{gapped} system, whose Hamiltonian is~\cite{Coleman:1975pw}:
\begin{align}
H_{\vf,D}  
&= \frac{1}{2} \!:\!\!\bigg( \Pi_\vf^2 + (\d_x \vf)^2 + g^2 \left(\vf + \frac{D_{\rm ext}}{g}\right)^2 \bigg)\!\!:~~,
\end{align}
where again $g^2 = e^2/\pi$, $D_{\rm ext}$ is the scalar potential that sources the external fields $F^{\mu \nu}_{\rm ext} = \ep^{\mu \nu} D_{\rm ext}$, $\Pi_\vf$ denotes the momentum canonically conjugate to the $\vf$-field, and :$(A...Z)$: denotes the normal ordered operators $A...Z$. As is discussed in detail in~\cite{Coleman:1975pw}, the on-shell excitations of the system, i.e. single-particle excitations of the $\vf$-field, are massive and correspond to neutral mesons, i.e. massive bound states of massless charged fermionic fields $\vf \sim \psi^+ \psi^-$. 

Following~\cite{Chu:2010xc}, we consider a configuration of external charges $\pm Q$ a distance of $L$ apart. The corresponding field strength tensor and the current are given by  
\begin{align}
&F^{\mu \nu}_{\rm ext} (x,t) \coloneqq \epsilon^{\mu \nu} D_{\rm ext}(x,t), \nonumber\\
&j^{\mu}_{\rm ext} \coloneqq \d_{\nu} F^{\mu \nu}_{\rm ext} = Q u^{\mu} \left( \delta\left(x+\tfrac{L}{2}\right) - \delta\left(x-\tfrac{L}{2}\right)\right)~\!.
\end{align}
where $D_{\rm ext}(x,t):=  Q\left[ \theta(x+L/2) - \theta(x-L/2)\right]$ with $\theta(x)$ is the usual step-function, $\theta(x) \coloneqq \int^x_{-\infty} dy\, \delta(y)$, and $u^{\mu} \coloneqq (1,0)$.
The equations of motion (EOMs) derived from $S_{\p,A}$ for $\p$ and $A_{\mu}$ (which is non-dynamical in $(1+1)$-dimensions) are given by
\begin{align}
{\rm EOMs}:~~\begin{cases}
~~~\d^2 \vf 		&\!\!\!\! =  g \ep^{\mu \nu}~ F_{\mu \nu}\,,\\
\d_\mu F^{\mu \nu} 	&\!\!\!\! = g \ep^{\mu \nu} ~\d_\nu \vf + j^{\nu}_{\rm ext}~~.
\end{cases}
\label{eqEOMi}
\end{align}
Crucially, these equations of motion are linear in the fields. We can thus straightforwardly integrate the non-dynamical $A^{\mu}$ field out of the equations of motion, which effectively reparametrizes the energy density stored in the electric field in terms of the value of the $\vf$-field at a point. In terms of this $\vf$-scalar field, the full space-time dependence of the E-field is given by~\cite{Chu:2010xc}
\begin{equation}
F^{\mu \nu} = \ep^{\mu \nu} (g \vf + F) + F_{\rm ext}^{\mu \nu} = \ep^{\mu \nu}\big( g \vf + D_{\rm ext} + F\big)~~,
\label{eqDF}
\end{equation}
where we have re-expressed the electromagnetic field-strength in terms of a single scalar potential, $D_{\rm ext}$, and $F$ is an additive integration-constant acquired when one moves from the EOM in Eq.~\eqref{eqEOMi} to the field-strength, and is further directly proportional to the $\theta$-term seen in the massive Schwinger model~\cite{Coleman:1976uz}: $\theta := 2 \pi F/e$.\footnote{Explicitly, this $F$ (or, equivalently, $\theta$)-offset in the relationship between the electromagnetic field-strength and the $\vf$-scalar meson field is:
\begin{align}
F^{\mu \nu} 
= \ep^{\mu \nu}\big( g \vf + D_{\rm ext} + F\big) 
= \ep^{\mu \nu} \bigg( g \bigg( \vf + \frac{\theta}{2 \sqrt{\pi}} \bigg) + D_{\rm ext}\bigg)~~.
\label{eqDFb}
\end{align}
For the massless Schwinger model, we may re-define $\vf \to \vf - \theta/(2 \sqrt{\pi})$ to eliminate this $\theta$-term. However, as we shall see in sections~\ref{secMassive1} and~\ref{secMassive2}, because finite fermion masses introduce $\cos(2 \sqrt{\pi} \vf)$-terms in the Hamiltonian, this shift of $\vf \to \vf - \theta/(2 \sqrt{\pi})$ induces a $\theta$-dependence into the cosine-potential in Eq.~\eqref{eqCosShift}~\cite{Coleman:1975pw, Coleman:1976uz}, via $\cos(2 \sqrt{\pi} \vf) \to \cos(2 \sqrt{\pi} \vf + \theta)$. } Rewriting $F^{\mu \nu}_{\rm ext}$ in this way will, later, allow us to explicitly map between the Hamiltonian-based and the action-based frameworks in, respectively, Refs.~\cite{Chu:2010xc} and~\cite{Coleman:1975pw}. 

Returning to capacitor discharge, we see that inserting $F_{01} = D_{\rm ext} + g \vf$ back into the equation of motion for $\vf$, we recover the well-known fact that the coupling between the non-dynamical electromagnetic field and the dynamical scalar field $\vf$, which does not have a mass-term within the action $S_{\vf,A}$, will source a mass-term for on-shell excitations of this scalar field. We therefore find
\begin{align}
\label{gmrel}
(\d^2 + m_\vf^2) \vf = - g D_{\rm ext} ~~,~~m_\vf^2 := g^2 = \frac{e^2}{\pi}~~.
\end{align}
In this way, we see explicitly that the bosonization map $:\bar \psi \g^{\mu} \psi: \leftrightarrow \d^\mu \vf$, which we used to naively bosonize the action for massless QED in 1$+1$ - dimensions reproduces the well-known gap in the Schwinger model~\cite{Schwinger:1962tp,Coleman:1975pw}. 

Because the equations of motion are linear in the $\vf$-basis, Ref.~\cite{Chu:2010xc} exploited linearity to solve for the dynamics of the scalar field profile due to the presence of a single charge $Q$ located at a definite position on the $x$-axis. The static solution $\vf(x)$ due to this single point-charge,
\begin{align}
(\d^2 + m_\vf^2) \vf = (g Q) ~{\rm sign}(x)~,
\end{align}
is, by using retarded Greens functions, found to simply be
\begin{align}
\vf_I(x) &= - g Q \int \frac{dk}{2\pi} \frac{\sin(kx)}{k(k^2 + m_\vf^2)} \nonumber\\
& = -\frac{1}{2}\frac{g Q}{m_\vf^2}~{\rm sign}(x)~\bigg( 1- e^{-m_\vf |x|} \bigg)~.
\label{eqVFI}
\end{align}
Note that the energy of this configuration diverges linearly with the system size $\lambda$, i.e. when $|x| \leq \la$ and $\la \gg L$, $\int_\lambda dx H_{\vf_I,D}(x) \sim \lambda Q^2 g^4/m_\vf^4$. This is another reflection of the fact that electric charges confine in (1+1)-dimensions, due to the linear growth of the scalar potential from a point-charge. 

Linearity of the fields themselves, and of the equations of motion, gives the total $\vf$ field-profile from the two initial capacitors with $\pm Q$ at $x = \pm L/2$ in terms of a difference of two such profiles:
\begin{align}
\vf_s(x) &= \vf_I\left(x-\tfrac{L}{2}\right) - \vf_I\left(x+\tfrac{L}{2}\right)   \nonumber\\
&= -\frac{gQ}{m_\vf^2} \times 
\begin{cases}
\qquad ~\! 
e^{-m_\vf |x|} \sinh(m_\vf L/2) 
\!\!\! &;~|x| > \tfrac{L}{2}~, \\
(1-e^{-m_\vf |L|/2} \cosh(m_\vf x)) 
\!\!\! &;~|x| < \tfrac{L}{2}~. 
\end{cases}
\label{eqVFS}
\end{align}
Note that these equal and opposite charges render the previously divergent energy of each individual point charge, now, finite. The total energy is sub-extensive in their separation.

The complete solution for the classical $\vf$-field profile that corresponds to the end-point of the capacitor's discharge, though, comes from imposing the initial-value condition that the only charged fermions at $t = 0$ comes from these two piles of charge $+Q$ and $-Q$, separated by a distance of $L$. Crucially, the initial conditions imposed by the starting capacitor configuration are:
\begin{align}
\vf(x, t)|_{t = 0} = \d_t \vf(x,t)|_{t = 0} = 0~.
\label{eqInitial}
\end{align}
At this point, we pause to note that the above initial-value condition is \emph{two}-fold. I.e., the full solution to the EOM, $\varphi(x,t)$ must have both a vanishing profile at $t = 0$ \emph{and} a vanishing time-derivative at $t = 0$. Crucially, as the static solution in Eq.~\eqref{eqVFS} is generically non-zero, it does not satisfy both conditions laid-out in  Eq.~\eqref{eqInitial}. Precisely because of this, we see that the system must dynamically evolve from a more complex $\varphi$-profile before finally relaxing to the late-stage, static profile in Eq.~\eqref{eqVFS}.

\begin{figure*}[t]
\includegraphics[width =\textwidth]{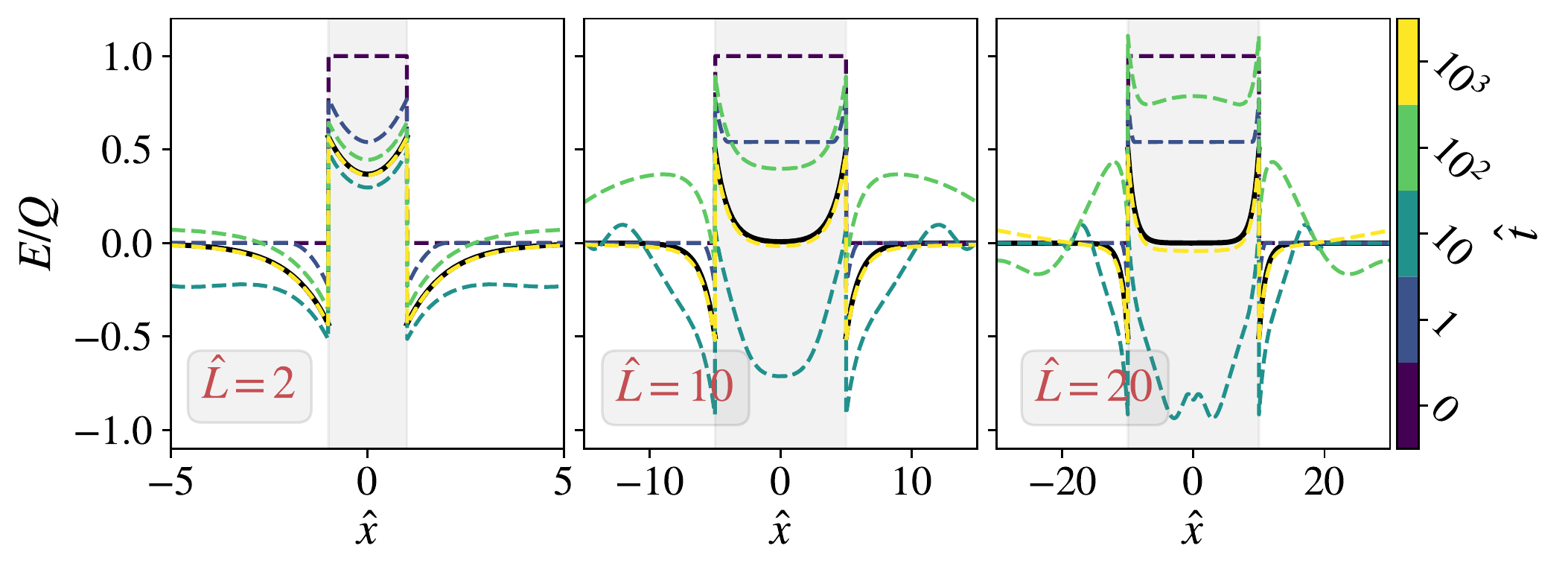}
\includegraphics[width=0.95\textwidth]{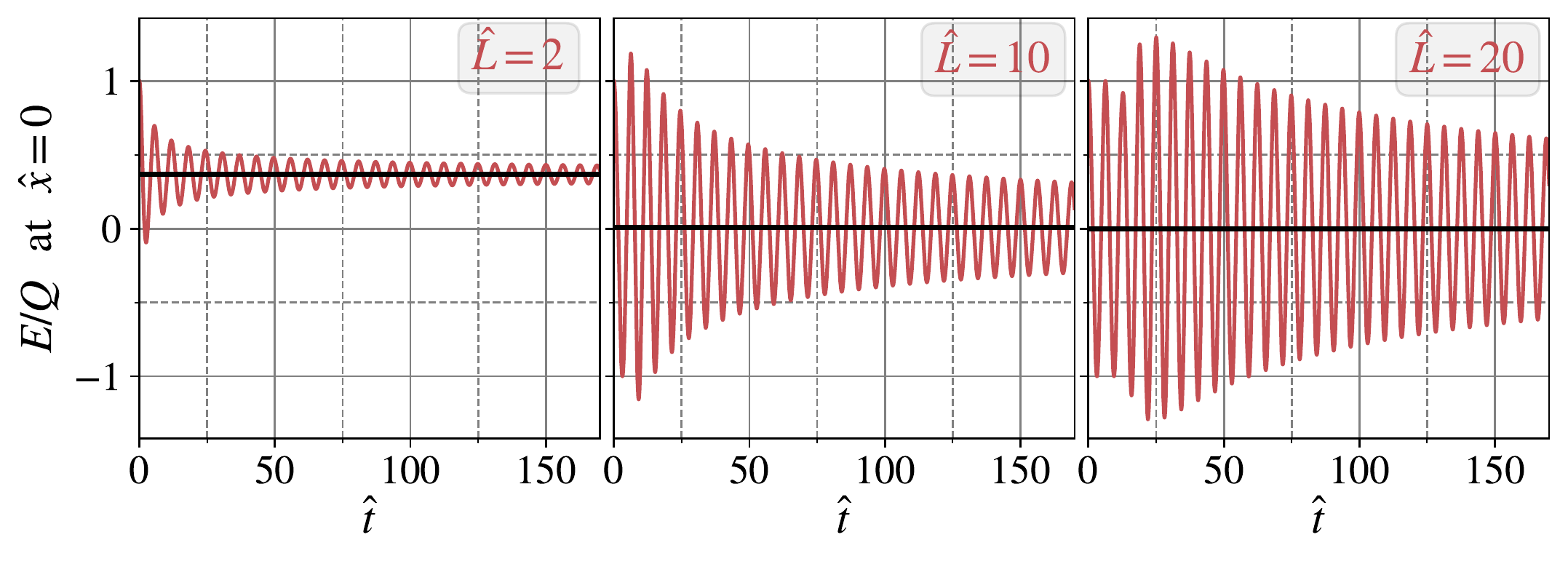}
\caption{Capacitor-discharge in \textit{massless} QED\t~via dipole discharge at plate separations $\hat{L} \equiv L/m_\vf = \{2, 10, 20\}$, at $x = 0$. The initial electric field discharges to the final field configuration (solid black line) undergoing oscillations along the way. Evidently, as one approaches the threshold separation at $L = m^{-1}_\p$, the final electric field never completely discharges within the capacitor.}
\label{Fig1}
\end{figure*}

It is straightforward to see that the total solution consistent with the initial conditions~\eqref{eqInitial} and the static final-state field configuration $\vf_s(x)$ in~\eqref{eqVFS} is simply,
\begin{align}
\vf(x,t) = - 2 g Q \int \frac{dk}{2\pi} \frac{\cos(kx) \sin(kL/2)}{k(k^2+m_\vf^2)} \bigg( 1 - \cos\left(t \sqrt{k^2 + m_\vf^2}~\right) \bigg)~.
\label{eqExact1}
\end{align}
These integrals are closely related to Bessel-J functions. For instance, as pointed-out in~\cite{Chu:2010xc}, we have
\begin{align}
\d_x \d_t \vf(x,t) &=  g Q \bigg[ J_0\left(m_\vf \sqrt{t^2 - x_-^2}~\right) \theta(t-x_-) 
- J_0\left(m_\vf \sqrt{t^2 - x_+^2}~\right) \theta(t-x_+) \bigg] ~,
\label{eqNice1}
\end{align}
where $x_\pm := x \pm L/2$, and the two terms correspond to the dynamical charges due to the left/right capacitor plates. We can find the late-time behavior of the electromagnetic decay-response prompted by the initial, unstable, charged capacitor conditions by manipulating these exact expressions. Fig.~\ref{Fig1} demonstrates the capacitor-discharge as a function of time $\hat{t} \equiv t/m_\varphi$ at $x = 0$ in massless $\mathrm{QED}_2$ for plate separations $\hat{L} \equiv L/m_\vf = \{2, 10, 20\}$. 
Immediately evident from both these figures and/or the related content in Eq.~\eqref{eqVFS}, is the appearance of a threshold plate separation below which discharge becomes inefficient. As we shall elaborate upon in the next section, this is consistent with our proposed physical interpretation of this process as being due to dipole discharge. 

\section{Physical interpretation of Schwinger production in (1+1)-$d$}\label{secMeson}

The goal of this investigation is to extend the analysis of~\cite{Chu:2010xc} to the case of massive QED\t. Before we do so, it is important to arrive at the correct physical interpretation of the results of~\cite{Chu:2010xc} reviewed in the previous section. We do this by clarifying the following four points: Firstly, we show that the Hamiltonian and Lagrangian formulations of Refs.~\cite{Chu:2010xc} and~\cite{Coleman:1975pw} match. This is slightly nontrivial, but of course had to happen. 

Secondly, we explicitly discuss how the two charges $\pm Q$, separated by a distance $L$, discharge via production of clouds of $\vf$-mesons. As we are in $d = 2$, electric charges confine. Thus the produced degrees of freedom cannot themselves be charged. However, they can and do have electric dipole moments. Thus the only way to discharge the initial capacitor is by producing clouds of polarized $\vf$ mesons. Producing such clouds requires clear threshold effects, somewhat different to the $e^{-\pi m^2/(q E)}$ threshold that appears in Schwinger's classic result. Indeed, there are three fundamental length-scales in this system, given by the separation between the capacitor plates $L$, the length-scale of the electric field, and the Compton wavelength set by the physical degrees of freedom of this system, i.e. the massive neutral $\vf$-mesons, $\ell_\vf$. This \textit{third} length-scale, which is ultimately due to the fact that QED is a relevant interaction in $d < 4$, is given in terms of the charge of the fundamental fermions of QED\t:
\begin{align}
m_\vf = \frac{e}{\sqrt{\pi}}~~,~~\ell_\vf = \frac{1}{m_\vf} = \frac{\sqrt{\pi}}{e}~~.
\end{align}
This characteristic length-scale gives an interesting wrinkle on how one may envision the initial conditions of highly-charged and well-separated capacitor plates---and for their subsequent (exact) time-evolution and eventual discharge. 

Thirdly, we note that these scales allow us to study capacitor discharge within a self consistent approximation in various limits. There are two independent dimensionless ratios of these length-scales. There is the ratio of the Compton wavelength to the separation between the capacitor plates, $L$, and thus 
\eq{}{L/\ell_\vf := q L = e L/\sqrt{\pi}.} 
Furthermore, there is the ratio of the Compton wavelength to the length-scale set by the electric field strength of the undischarged capacitor, 
\eq{}{\ell_F := 1/Q,} 
and so $\ell_F/\ell_\vf = 1/(Q \ell_\vf) = m_\vf/Q = q/Q$. 

Fourth and finally, we would like to discuss the extent to which solutions to the classical equations of motion capture the full quantum mechanical evolution of capacitor discharge in massless QED\t. That is, when do the solutions for the classical equations of motion solve for the full quantum dynamics of a given system? We address each of these points in the following subsections. 

\subsection{Actions and Hamiltonians}

To begin with, Ref.~\cite{Chu:2010xc}'s computations were done using the equations of motion derived from the action $S_{\vf,A}$. When one evaluates the equations of motion for the naively massless scalar field $\vf$, one sees that integrating-out the non-dynamical electromagnetic field induces a non-zero mass, $m_\vf^2 = e^2/\pi$. As Ref.~\cite{Chu:2010xc} notes, this is consistent with the gap in the Schwinger model~\cite{Schwinger:1962tp}. Yet, as the field content in $S_{\vf,F}$ and $H_{\vf,D}$ seem to be fundamentally different, one may worry whether the descriptions match exactly. To see that they do, recall the expression for the 00-component of the stress-energy tensor derived from the action, $S_{\vf,F} \mapsto T^{\mu \nu} := \delta S_{\vf,F}/\delta g_{\mu \nu}$, in~\cite{Chu:2010xc}:
\begin{align}
T^{00} = \frac{1}{2} \phantom{}:\!\! \bigg( \Pi_\vf^2 + (\d_x \vf)^2 + m_\vf^2\left( \vf + \frac{D_{\rm ext}}{m_\vf}\right)^2 \bigg)\!\!:
\end{align}
Here we made free use of the normal ordering prescription as these two functionals are implicitly, for us, evaluated in reference to specific field profiles for $\vf(x,t)$ and $D_{\rm ext}(x,t)$ given in section~\ref{secMassless}. 

Thus, we see that the local expression for the 00-component of $T^{\mu \nu}$ derived from $S_{\vf,F}$ from~\cite{Chu:2010xc} and the Hamiltonian $H_{\vf,D}$ from~\cite{Coleman:1975pw}, when evaluated on specific, shared, profiles for both $\vf$ and $\ep^{\mu \nu} D_{\rm ext} = F^{\mu \nu}_{\rm ext}$,  indeed match exactly: 
\begin{align}
T^{00}(x,t) \!=\! H_{\vf,D}(x,t)~.
\end{align}
In section~\ref{secMassive2}, we use this and the explicit results for the \textit{massive} Schwinger model's Hamiltonian in (1+1)-dimensions in~\cite{Coleman:1975pw} to extend the results of Ref.~\cite{Chu:2010xc} to capacitor discharge in \textit{massive} QED\t. We do so by first studying how the various length-scales in the problem interplay to determine the late-time, discharged, field profile. We then study the time-scales for this dynamical relaxation.

\subsection{Capacitor discharge and $\vf$-meson clouds: Length-scales}\label{secClouds}

The late-time, static, E-field profile is the sum of two terms, $E_{\rm static}(x) := D_{\rm ext} + g \vf_s(x)$, which are respectively given in Eqs.~\eqref{eqDF} and~\eqref{eqVFS}. It takes the following explicit form:
\begin{align}
F_{01}(x,t)\bigg|_{t \to \infty} \!\!\!\!\!\!\!\!\!\! = E_{\rm static}(x) = 
\begin{cases}
-\frac{Q}{2} \times \left({\rm Exp}\left(-\tfrac{2|x|-L}{2\ell_\vf}\right) - {\rm Exp}\left(-\tfrac{2|x|+L}{2\ell_\vf}\right)\right), &\!\! |x| > L/2~, \\
+\frac{Q}{2} \times \left(1 - {\rm Exp}\left(-\tfrac{L-2|x|}{2\ell_\vf} \right) +1- {\rm Exp}\left(-\tfrac{L+2|x|}{2\ell_\vf}\right) ~\right), & \!\! |x| < L/2~.
\end{cases}
\label{eqLateE}
\end{align}
The energy stored in the electric field between the two capacitor plates, $\hhh(t)$, is given by
\begin{align}
\hhh(t) := \frac{1}{2} \int_{-L/2}^{L/2} dx \bigg( F_{01}(x,t)^2 \bigg) = \frac{1}{2} \int_{-L/2}^{L/2} dx \bigg( D_{\rm ext}(x) + g \vf(x,t) \bigg)^2~,
\label{eqHHH}
\end{align}
where $\vf(x,t)$ is as given in~\eqref{eqExact1} and $D_{\rm ext}(x)$ is, again, the field profile in Eq.~\eqref{eqDF}.

Now, to our first point: How the characteristic length-scale for the size of the $\vf$-meson enters into threshold effects associated with capacitor discharge. One of the most clear instances of this comes from the energy stored in the electric field \textit{after} the discharge has completed, at $t \to \infty$. Noting that the transient solutions eventually damp-out to the static solution, $\vf_s(x)$ in Eq.~\eqref{eqVFS}, we can compute the energy that remains in the capacitor plate geometry after discharge to be~\cite{Chu:2010xc},
\begin{align}
\begin{cases}
\hhh(0) ~= \int_{-L/2}^{L/2} dx \big(T^{00}(x,t)\big|_{t \to 0}\big) ~~\!\!= \frac{Q^2 L}{2} ~,
\\
\hhh(\infty) = \int_{-L/2}^{L/2} dx \big(T^{00}(x,t)\big|_{t \to \infty} \big) \!= \frac{Q^2 L}{2} \left( \frac{ 1 - e^{-2 L m_\vf}}{2 L m_\vf} \right) = \hhh(0) \times \bigg(1 - \tfrac{L}{\ell_\vf} + {\cal O}\big( \tfrac{L^2}{\ell_\vf^2} \big) \bigg) ~.
\end{cases}
\label{eqLateH}
\end{align}
This result for the residual energy stored in the E-field of the fully discharged capacitor has several interesting features. Note that this shows us that the final-state energy of the discharged capacitor is \textit{sub-extensive} in the the ratio between the separation of the two capacitor plates and the length-scale associated with producing a \textit{pair} of neutral $\vf$-mesons. Running this exponential suppression in the other direction we see that this guides our interpretation of \textit{how} a capacitor could discharge in massless QED\t~at all: Namely, capacitor discharge can only happen via emission/creation of a \textit{pair} of (clouds of) $\vf$-mesons whose electric dipole moments are aligned to cancel and screen the dipole generated by the external field itself, in effect screening the original charges of $\pm Q$.

Importantly, if the separation between these capacitor plates is insufficient to physically \textit{fit} any such mesons between them, then the external fields set by the capacitor cannot adequately discharge. Furthermore, we expect the discharge to become inefficient as the plate separation $L$ approaches $\ell_\vf$, as evident from the leftmost panel in Fig.~\ref{Fig1}. Indeed, whenever $L \lesssim \ell_\vf$, then it is simply not possible to create any static capacitor-plate geometry out of a configuration of these neutral $\vf$-mesons within QED\t. (We comment further on the feasibility of creating a capacitor of charges $\pm Q$, separated by a distance $L$, \textit{directly} out of these $\vf$-mesons---in a manner analogous to how actual capacitor plates are constructed with actual electrons deposited on physical metal plates in a transistor radio ---in our concluding remarks in section~\ref{secEnd}.) 
Furthermore, the final electric field configuration in Eq.~\eqref{eqLateE} has exactly the interpretation of a pair of clouds of neutral $\vf$-mesons screening each capacitor plate of charge $\pm Q$. The difference between the magnitude of the net electric field across each plate is given by $\pm Q$ while the sum of the two fields just across each plate is $Q e^{-L/\ell_\vf}$. The net polarization of the $\vf$-meson fields acts to \textit{dipole screen} the initial charges $\pm Q$, and the penetration-depth of each cloud is given by the Compton wavelength of the $\vf$-mesons themselves: $\lambda_{\rm depth} = \ell_\vf$. 

\subsection{Capacitor discharge and $\vf$-meson clouds: Time-scales and late time tails}

Next, we turn to the important question of timescales associated with producing the $\vf$-meson clouds that discharge the electric field (see also \cite{Kukuljan:2021ddg} for a more detailed study). In this context, the evolution of the energy stored in the E-field between the two plates, $\hhh(t)$, plays a central role. We identify three time-scales: One associated with the speed of the initial capacitor discharge at $t = 0$ which we call $\tz$, and two associated with the two time-varying terms in the expression for $\hhh(t)$ itself which we denote $\tvv$ and $\tvs$. Before we determine these scales, we again stress that although massless QED\t~has $g = m_\vf$ (cf. eq (\ref{gmrel})), this degeneracy will turn out to be broken in massive QED\t. For this reason, in our analysis of the relevant timescales, we carefully distinguish between factors of $g$ and factors of $m_\vf$.

We define the first timescale $\tz$, associated with instantaneous capacitor discharge at $t = 0$:
\begin{align}
	\frac{1}{\tz^2} := \frac{\hhh''(t)}{\hhh(t)} \bigg|_{t \to 0}~.
\end{align}  
The time-variation within $F_{01}(x,t)$ is isolated in the cosine-term in Eq.~\eqref{eqExact1}, we have
\begin{align}
F_{01}(x,t) = E_{\rm static}(x) + E_{\rm vary}(x,t),
\end{align}
where
\begin{align}
E_{\rm vary}(x,t) = Q g^2 \!\int \frac{dk}{2\pi} \frac{\cos(kx) \sin(kL/2)}{k(k^2 + m_\vf^2)}\cos(t \sqrt{k^2 + m_\vf^2}).
\end{align}

It is straightforward to see that $\hhh''(0) = g^2 Q^2 L/2$, while $\hhh(0) = Q^2L/2$, and thus
\begin{align}
\frac{1}{\tz^2} = g^2 \implies \tz = \frac{1}{g} = \frac{\sqrt{\pi}}{e}~.
\label{eqTZ}
\end{align}
Now, the timescale $\tz$ characterizes the \textit{initial} rate of capacitor discharge. Further, it depends on the charge of the electron in QED\t, but does not depend on the mass of the $\vf$-boson itself. Thus, this timescale for capacitor discharge is common to both massless and massive ``electrons'' of charge $e = g \sqrt{\pi}$. Put differently, the initial rate of discharge of the capacitor's E-field is independent of the mass, $m_\psi$, for the degrees-of-freedom in the Lagrangian for either massless or massive QED\t.\footnote{We can qualitatively and quantitatively tie this independence to the fact that, as highlighted by the expressions for $\hhh(t)$ in Eq.~\eqref{eqHHH} and by the expressions/initial conditions for $\vf(x,t)$ in Eqs.~\eqref{eqInitial} and~\eqref{eqExact1}, all aspects of the $\vf$-boson's mass explicitly drop out of the expressions for $\hhh(t)$ and its time derivatives, when evaluated at $t = 0$.}

The other two important timescales we derive come from the late-time behavior of $\hhh(t)$, e.g. when $t \gg |x \pm \tfrac{L}{2}|, 1/m_\vf, m_\vf L^2$. As pointed out in \cite{Chu:2010xc}, the spatial current can be evaluated as
\begin{align}
j^x(x,t) 
&= g \d_t \vf(x,t) = - 2 g^2 Q \int \frac{dk}{2\pi} \frac{\cos(kx) \sin(kL/2)}{k(k^2+m_\vf^2)^{1/2}} \sin(t (k^2 + m_\vf^2)^{1/2}) \\
&= -\frac{g^2 Q}{2} \int_0^{L/2} d\ell \bigg[ J_0\left(m_\vf\sqrt{t^2 - x_{\ell-}^2}\right) \theta(t-x_{\ell+}) + J_0\left(m_\vf\sqrt{t^2 - x_{\ell+}^2}\right) \theta(t-x_{\ell+}) \bigg]~, \!\!\!\! 
\end{align}
where, here, $x_{\ell\pm} := x \pm \tfrac{L}{2}$. At late times $t \gg |x \pm L/2|$, this expression can be expanded in terms of $J_0(m_\vf t)$-type Bessel functions which, when $t \gg 1/m_\vf, m_\vf L^2$, can be further expanded as
\begin{align}
&
j^x(x,t) \bigg|_{t \gg \tfrac{1}{m_\vf}, ~m_\vf L^2, ~|x_\pm|} = 
- \frac{Q g^2 L}{\sqrt{2 \pi m_\vf t}} \bigg( \cos\left(m_\vf t - \frac{\pi}{4} \right) + {\cal O}\left( \frac{m_\vf L^2}{t} \right) \bigg)~.
\end{align}
Since $\d_t F^{tx}(x,t) = \d_t E(x,t) = j^x(x,t)$, for the electric field we obtain
\begin{align}
E(x,t) = E_{\rm static}(x) + \int dt \big( j^x(x,t) \big)~.
\end{align}
Thus, at late times, the E-field between the capacitor plates approaches
\begin{align}
\!\!\! 
E_\mathrm{vary}(t)\bigg|_{t \gg t_\star} \!\!\!\!\!\!\!\!\!\!
=  - \!\! \int dt \frac{Q g^2 L}{\sqrt{2 \pi m_\vf t}} \bigg( \cos\left(m_\vf t - \frac{\pi}{4} \right) + {\cal O}\left( \frac{m_\vf L^2}{t} \right) \bigg)  ,
\!\!\!\!
\end{align}
where $t \gg t_\star$ is a proxy for the timescales such as $|x_\pm|, 1/m_\vf, m_\vf L^2$. This implicitly defines $E_{\rm vary}(t)$ at late times. 

Using this late-time expression, we see that at late times the electric field between the two capacitor plates, i.e. for $|x| \leq L/2$, is the sum of a term that varies in space but not in time and a term that varies in time but not in space. Consequently, the energy stored between the capacitor plates, $\hhh(t)$, is the sum of three terms:
\begin{align}
\hhh(t)|_{t \gg t_\star} = \frac{1}{2} \int_{-L/2}^{L/2} dx \bigg\{ E_{\rm vary}(t) + E_{\rm static}(x) \bigg\}^2~,
\end{align}
and, again, $t \gg t_\star$ is a proxy for the various timescales discussed above. As the oscillatory envelope of $E_{\rm vary}(t)$ decays as $1/\sqrt{t}$, at extremely late times only the spatially varying term matters, i.e. when $t \to \infty$. 

We are interested in the characteristic time scales of the decay of the second two terms in $\hhh(t)$. To render the expressions appropriately dimensionless (and to follow~\cite{Chu:2010xc}), we consider the ratio
\begin{align}
\frac{\hhh(t)}{\hhh(0)} \bigg|_{t \gg t_\star}^{\rm RMS} := \frac{T_1}{t} + \frac{T_2}{\sqrt{t}} + T_3 ~,
\end{align}
where ``RMS'' refers to replacing the oscillatory parts of $E_{\rm vary}(t)$ with their root-mean-square averages. Now, the term $T_3$ is recognizable as the ratio of the energy stored in the final-state of the capacitor to the energy stored in the initial state, which was computed in Eq.~\eqref{eqLateH}.  

The first two terms, $T_1/t$ and $T_2/\sqrt{t}$, strike at the heart of this late-time tail of the discharge. They define characteristic time-scales associated to the discharge of the capacitor which we denote $\tvv$ and $\tvs$, respectively. Explicitly, these are defined as:
\begin{align}
\tvv := T_1
~~, ~~
\tvs := T_2^2
~~.
\end{align}
After straightforward computations, and carefully distinguishing between $g$ and $m_\vf$, we find
\begin{align}
	\bigg(
	\frac{T_1}{t}
	, 
	\frac{T_2}{\sqrt{t}}
	\bigg) 
	\mapsto (
	\tvv
	,
	\tvs
	) 
	= \bigg(
	\frac{1}{g}~ \frac{g^2 L^2}{4 \pi}  \bigg(\frac{g}{m_\vf}\bigg)^3
	~,~
	\frac{1}{m_\vf} ~\frac{Q^2 L^2}{\pi} \bigg(\frac{g}{m_\vf}\bigg)^6 \bigg(1 - \frac{1-e^{-m_\vf L}}{m_\vf L}\bigg)^2
	\bigg)
	\label{eqTVV}
\end{align}
Now, because $g = m_\vf$ for massless QED\t, we see that $\tvv$ matches the timescale Ref.~\cite{Chu:2010xc} associated with capacitor discharge. Crucially, $\tvs$ represents a new timescale for the tail of the discharge: 
\begin{align}
( 
\tvv 
, 
\tvs
) \bigg|_{g \to m_\vf} = 
\bigg( 
\frac{1}{g} ~\frac{g^2 L^2}{4 \pi}
~~,~~ 
\frac{1}{g} ~\frac{Q^2L^2}{\pi} \bigg(1 - \frac{1-e^{ -g L}}{g L}\bigg)^2
\bigg)
\end{align}
Note that between the two terms in the expression for the decay of the energy stored in the capacitor with respect to time, $T_2/\sqrt{t}$ and $T_1/t$, the $T_2$-term dominates at late-times, $t \gg t_\star$. (Further, because we are at late-times, \textit{both} terms should be small.) Thus the timescale $\tvs$ plays a vital role in capacitor discharge dynamics. There are several interesting features of these timescales. 

First, as noted above, the initial rate of capacitor discharge set by $\tz$ is independent of all details of the system save for the charge of the fermions in the QED\t~action. This is physically sensible. At the very onset of the discharge, the response of the system to any applied field should be governed by how the fundamental, short-distance, degrees of freedom couple to this field. It does not have time to ``see'' the size of the system. Fundamentally, $\hhh''(0)/\hhh(0)$ equals $g^2$ because the $\vf$-mesons themselves source electric fields via, effectively, an overall factor of $g^2$.

Second, the two timescales that dictate how energy stored in the discharging capacitor dissipates are each (super)-extensive in terms of the physical size of the capacitor. Put differently, each timescale is proportional to $L^2$: $\tvs \propto L^2$ and $\tvv \propto L^2$. This seems sensible as the final energy in the capacitor is \textit{sub}-extensive in $L$. Further, the latest-time tails in the energy stored in the capacitor are governed by the $\sqrt{\tvs}/\sqrt{t} = T_2/\sqrt{t}$-term in the RMS-averaged ratio $\hhh(t)/\hhh(0)$. Crucially, this timescale is also proportional to the initial energy stored in the capacitor, namely $\tvs \propto Q^2$.

\subsection{Quantum evolution from classical equations of motion?}\label{sDynamics}

Finally, we turn to the question of when knowledge of the classical dynamics of a system is sufficient to capture the full quantum dynamics. There are three possibilities. For a free theory, one can construct coherent states that form an over-complete basis of states. Any expectation value of field operators in a coherent state will satisfy the classical equations of motion. However, whether the specific initial conditions of charged capacitor plates can be realized this way is a non-trivial question. A second possibility also presents itself when the theory is integrable (in other words, fully solvable), as for example is the case for the sine-Gordon model in (1+1)-$d$. The latter would be necessary were we interested in the full range of the dynamics of the system, particularly in states with low occupation numbers, and is an interesting exercise in principle worthy of a followup investigation that we will not pursue here. 

Instead, we focus on the third and simplest possibility -- that one can work within a self consistent approximation in the context of initial states with large occupation numbers, a correspondence can also be shown to hold in out of equilibrium and interacting systems \cite{Mueller:2002gd}. In the present context, this requires us to work in the limit that the charge of the capacitors be vastly larger than the fundamental quanta of charge. That is:
\begin{align}
	N:= \frac{Q}{e} \gg 1.
\end{align}
In this limit, we can be sure to recover the fully quantum evolution of the discharging capacitor from the classical equations of motion even if the classical equations of motion are non-linear. To see this in detail, we adapt the treatment reviewed in \cite{Mueller:2002gd} and consider a generic scalar field theory with canonical kinetic term and arbitrary potential. 

Consider the finite time correlation functions of some observable $\calO(t)$ consisting of a string of $\p$ operators evaluated at time $t$ (but not necessarily at spatially coincident points)
\eq{ftcf}{\langle \calO(t)\rangle=\int \mathcal{D}[\p_t] \calO[\p_t] \int \mathcal{D}\left[\p_{0}\right] \rho\left[\p_{0}\right] U^{*}\left[\p_t, \p_{0}, t-t_{0}\right] U\left[\p_t, \p_{0}, t-t_{0}\right],}
where $\rho[\p_0]$ is the initial density matrix corresponding to some initial field profile $\p(t_0,x) := \p_0$, and where
\eq{}{U\left[\p_t, \p_{0}, t-t_{0}\right]=\int \mathcal{D}[\p(t')] e^{i \int_{t_{0}}^{t} L[\phi(t')] d t'}}
integrates over all field profiles that interpolate between $\p_0$ and $\p(t,x) := \p_t$. 
Instead of viewing $U$ and $U^*$ as containing separate functional integrals over the same field $\p$, a standard trick employed in real-time perturbation theory is to view the fields in the separate integrations as being distinct fields, $\p_-$ and $\p_+$, and writing the Lagrangian as
\eq{lag}{ L = \int dx \left[\frac{1}{2}\left(\partial_{\mu} \p_{-}\right)^{2}-\frac{1}{2} m^{2} \p_{-}^{2}-\sum_{n=3} \frac{c_n}{n !} \p_{-}^{n}\right]-   \int dx \left[\frac{1}{2}\left(\partial_{\mu} \p_{+}\right)^{2}-\frac{1}{2} m^{2} \p_{+}^{2}- \sum_{n=3}\frac{c_n}{n !} \p_{+}^{n}\right]}
so that
\eq{}{U^{*} U=\int \mathcal{D}\left[\p_{-}\right] \mathcal{D}\left[\p_{+}\right] e^{i \int_{t_{0}}^{t} L\left[\p_{-}, \p_{+}\right] d t'},}
where $\p_-, \p_+$ are both equal to $\p(t,x)$ and $\p(t_0,x)$ at $t$ and $t_0$ respectively. Making the field redefinitions (where the meaning of the subscripts will become clear shortly)
\eq{}{\p_c := \frac{1}{2}\left(\p_- + \p_+\right),~~~ \p_q := \p_- - \p_+,}
and working with systems where both $\p_+$ and $\p_-$ correspond to field profiles with large initial occupation, and so are themselves large, one can show under certain assumptions on $\rho[\p_0]$ that $\p_c$ can also naturally be taken to be large whereas $\p_q$ is small \cite{Mueller:2002gd}. Therefore, rewriting the action (\ref{lag}) in terms of $\p_c$ and $\p_q$ and expanding to first order in $\p_q$ one finds
\eq{}{L[\p_c,\p_q] =\int dx \left[ \partial^{\mu} \p_c \partial_{\mu} \p_q-m^{2} \p_c \p_q- \p_q\sum_{n=4}\frac{c_n \p_c^{n-1}}{(n-1) !}\right] + ...}
where the ellipses denote terms of order $\p_q^2$ and higher. The final term in the square parentheses is simply the Taylor expansion of $\p_q V'(\p_c)$. Recalling that our boundary conditions are such that $\p_q \equiv 0$ at $t$ and $t_0$ and $\p_t = \p_c(t,x)$ and $\p_0 = \p_c(t_0,x)$, so that Eq. \eqref{ftcf} becomes
\eq{lc}{\langle \calO(t)\rangle=\int \mathcal{D}[\p_c(t)] \calO[\p_c(t)] \int \mathcal{D}\left[\p_c(t_0)\right] \rho\left[\p_c (t_0)\right] \int \mathcal{D}\left[\p_{c}\right(t')] \mathcal{D}\left[\p_{q}\right(t')] e^{i \int_{t_{0}}^{t} L\left[\p_{c}, \p_{q}\right] d t'}.}
Because the assumption of large initial occupation forces $\p_q$ to be small enough so that higher order terms in the expansion Eq. \eqref{lc} can be neglected, we see that its functional integral straightforwardly enforces the constraint on the measure $\mathcal D[\p_c(t')]$:
\eq{}{\int \mathcal{D}\left[\p_{c}\right(t')] \mathcal{D}\left[\p_{q}\right(t')] e^{i \int_{t_{0}}^{t} L\left[\p_{c}, \p_{q}\right] d t'} = \int \mathcal{D}\left[\p_{c}\right(t')] \prod_{x, t'} 2 \pi \delta\left[\left(\square+m^{2}\right) \p_c({x}, t')+ V'[\p_c({x}, t')]\right],}  
meaning that only fields that satisfy the classical equations of motion,
\eq{ceom}{\left(\square+m^{2}\right) \p_c + V'[\p_c] = 0}
contribute to the functional integral in Eq. \eqref{lc}. Therefore, characterizing the full quantum  of the system in the limit of large occupation number boils down to knowledge of the classical solution to the equations of motion Eq. \eqref{ceom}. Defining the kernel
\eq{}{K\left[\p_t, \p_0, t-t_{0}\right]:=\int \mathcal{D}[\p_c(t')] \mathcal{D}[\p_q(t')] e^{i \int_{t_{0}}^{t} L[\p_c, \p_q] d t'},}
we see that it is annihilated by the classical equations of motion, and at the initial time satisfies 
\eq{}{\lim_{t \to t_0} K[\p(t),\p(t_0),t-t_0] = \prod_{x}\delta[\p(x,0) - \p_0(x,0)],}
which manifests the classical nature of the system on all expectation values
\eq{}{\langle \calO(t)\rangle=\int \mathcal{D}[\p_t] \mathcal{D}\left[\p_{0}\right] \calO[\p_t] K\left[\p_t, \p_{0}, t-t_{0}\right] \rho\left[\p_{0}\right].}
A direct corollary of the above is that an initial density matrix sharply peaked around the profile $\p_0(t_0,x)$ returns the expectation value $\langle \p^n(t,x) \rangle = \p_c^n(t,x)$, with $\p_c$ satisfying Eq. \eqref{ceom} with initial conditions $\p_c(t_0,x) \equiv \p_0(t_0,x)$.

\section{Bare fermion masses in the Schwinger model}\label{secMassive1}

We now turn our attention towards the addition of bare fermion masses to the Schwinger model. When the bare mass of the fermion in QED\t~vanishes, the mass of the neutral pion equals the mass-scale set by the effective coupling constant $g$ in the term $g \vf \ep^{\mu \nu} F_{\mu \nu}$ within the action $S_{\vf,A}$, i.e. $g = e/\sqrt{\pi}= m_\vf $. However, when this fermion has a bare mass, $m_\psi \neq 0$, the spectrum of stable excitations in the theory becomes much more rich~\cite{Coleman:1976uz}. That is, introducing a bare fermion mass distinguishes between the two mass-scales $g$ and $m_\vf$ that characterize the physical excitations of massive QED\t.

As discussed in Refs.~\cite{Coleman:1975pw, Coleman:1976uz}, when the fermions in QED\t~acquire a bare mass, the Hamiltonian for the dual $\vf$-meson field, is modified in the following way:
\begin{align}
H_{\vf,D} \to H_{\vf,D}(m_\psi)  
&= H_{\vf,D} + \Delta H_{\vf,D} ~
\end{align}
where
\begin{align}
\Delta H_{\vf,D} \equiv m_\psi e C :\!\cos\left(2 \sqrt\pi \vf + \theta \right)\!:.
\label{eqCosShift}
\end{align}
Here $C = e^{\gamma_E}/(2 \pi^{3/2}) \simeq 0.16$ is a numerical constant~\cite{Gross:1995bp}, $m_\psi$ is the bare mass of the electron~\cite{Coleman:1975pw}, and $\theta$ is a non-dynamical real parameter that lies within the range $\theta \in [-\pi, \pi]$ that, ultimately, originates from the derivative acting on $\vf$ in the duality/mapping $J^\mu \leftrightarrow \ep^{\mu \nu} \d_\nu \vf$ (as seen in e.g. Eq.~\eqref{eqEOMi}). It is somewhat striking that this new Hamiltonian explicitly depends on the parameter $\theta$ that characterizes the one-parameter family of vacua of the Schwinger model~\cite{Coleman:1975pw}. Crucially, when $m_\psi = 0$, the vacua are degenerate. However, we see that when the fermion has a bare mass within the QED\t~action, this degeneracy is lifted. Equivalently, the action for the dual scalar is,
\begin{align}
S_{\vf,D}  \to S_{\vf,D}(m_\psi) = S_{\vf,D} +\Delta S_{\vf,D} ~.
\end{align}
where
\begin{align}
\Delta S_{\vf,D} \equiv -C \bigg( \int d^2 x ~(m_\psi e) :\!\cos\left(2 \sqrt\pi \vf + \theta \right)\!: \bigg)~.
\end{align}

At this point, one could attempt to use the fact that for small values of $\vf$ the cosine-potential, $\cos(2 \sqrt{\pi} \vf)$, the equation of motion is approximately linear. In this context, the equations of motion that dictate capacitor discharge in both massive and massless QED\t~would continue to be linear and can straightforwardly be shown to be solved by a similar integral to Eq.~\eqref{eqExact1}, where now 
\eq{mm}{m_\vf = \sqrt{g(g + 2 e^{\gamma_E} m_\psi )},~~~ g = \sqrt{e^2/\pi}~.} Self-consistency of this approximation would thus require that $\vf(x,t)$, which solves the \textit{linearized} equation of motion, remains bounded and small at all times; if $\vf(x,t)$ is ever not-small, then we must consider the non-linear higher-order terms in the cosine-potential. Put differently, even though the two scales $g$ and $m_\vf$ are split when the fermion has a bare mass, $m_\psi \neq 0$, if it is consistent to both fix $\theta \mapsto 0$ and to remain in the small-$\vf$ regime, then capacitor discharge in massive QED\t~would still be described by the dynamics found in~\cite{Chu:2010xc}.

However, there can be no appreciable capacitor discharge in this limit. The reason is as follows. As commented in section~\ref{sDynamics}, we cannot trust the equation of motion to describe capacitor discharge unless we can guarantee large-occupation numbers~\cite{Mueller:2002gd}. In this context, we must have $N \gg 1$. Now, note that $\vf$ scales as $\vf \sim - 2 (g^2/m_\vf^2) N \times I_L(x,t)$, where $I_L(x,t)$ is an integral that depends only on the separation between the capacitor plates, $L$, and on $x$ and $t$. (Crucially, $I_L(x,t)$ does not depend on $N$.) When written in this way, we see that the range of capacitor configurations where $|\vf| \ll 1$ holds throughout the discharge is dramatically restricted at large-$N$.
Further, as the electric field evolves as
\begin{align}
E(x,t) 
&= D_\mathrm{ext} - g\vf(x,t) = D_\mathrm{ext} - \frac{Q}{N}\vf(x,t)~, 
\label{eqNice00}
\end{align}
we see that even if $|\vf|$ is bounded by a fixed small number (i.e. $|\vf| \ll1$), then the discharge is a $1/N$ effect. 
Indeed, after all is said and done, it is possible to show that $\hhh(0) - \hhh(\infty) \lesssim m_\vf$ in this scenario. 
In the next section, we study the numerical solutions to the full non-linear equations of motion for this massive bosonic model, which resembles a mass-deformed sine-Gordon model. This analysis is at once more general, in the sense that it only requires the large occupation approximation so that integrating the classical equations of motion is equivalent to solving for the full quantum dynamics of the system, and less analytically tractable. That said, the numerics speak for themselves.

\section{Integrability of capacitor discharge in massive QED\t}\label{secMassive2}

\begin{figure*}[t]
\includegraphics[width = \textwidth]{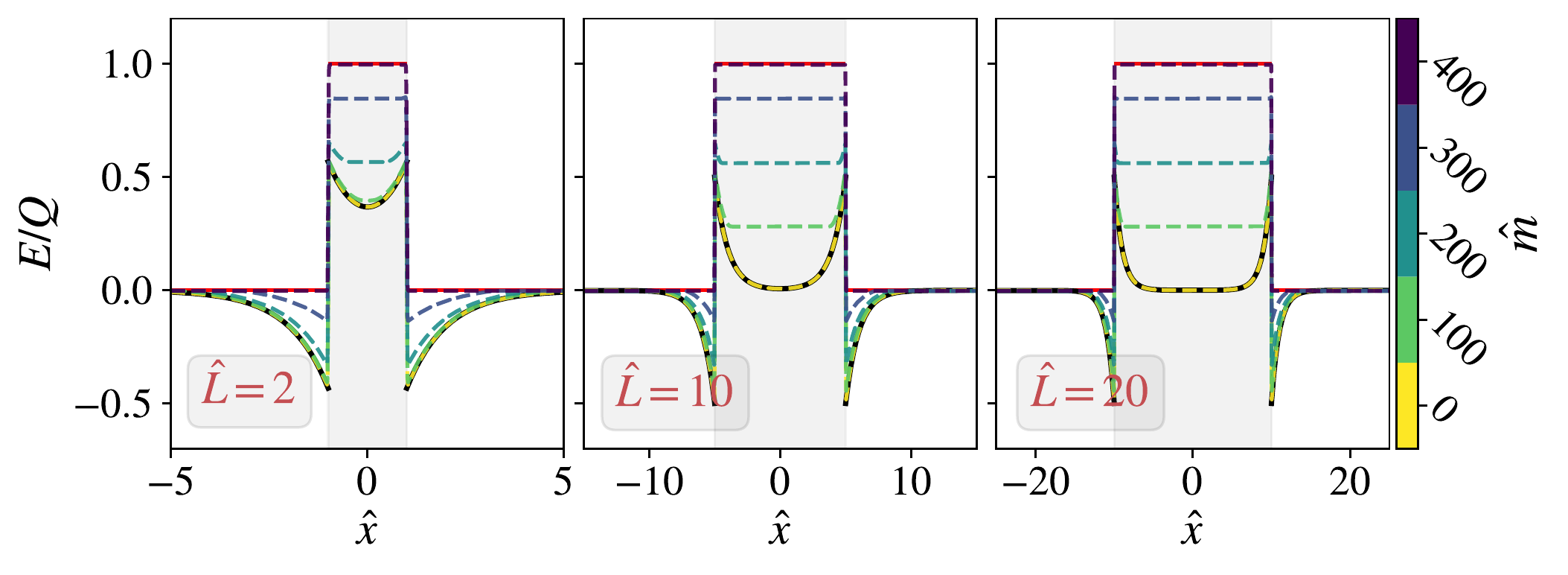}
\caption{
Numerically obtained asymptotic electric field profiles at the end of capacitor discharge in \textit{massive} QED\t~for three different plate separations (corresponding to separate panels), and for a range of mass parameter $\hat{m}$ (corresponding to the color-coded dashed curved), with $\hat{N} = 200$. The electric field profile prior to discharge is shown in solid red line. The black solid lines underneath the yellow dashed curves correspond to the analytical result Eq.~\eqref{eqLateE} in the massless situation, which agrees with our numerical solution. The existence of the mass threshold is visible in the fact that, at finite bare fermion mass ($g < m_\p$), the electric field never completely discharges.}
\label{Fig2}
\end{figure*}

As discussed above, in the limit where $N = Q/e$ is large, we are in a regime where the system is at large initial occupation, which as detailed in Sec. \ref{sDynamics}, implies that the expectation value of the operator equation of motion is equivalent to solving the classical equation of motion
\begin{align}
\label{bseom}
(\partial^2 + g^2) \vf_c = gD_\mathrm{ext} +  \bigg( \frac{e^{\gamma_E}}{\pi} ~m_\psi g  \bigg) \sin(2 \sqrt{\pi} \vf_c + \theta).
\end{align} 
That is, the quantum dynamics of capacitor discharge is completely characterized by solutions to the above equation of motion for large enough initial charge. In order to facilitate numerical integration, we consider the dimensionless variables,
\begin{align}
\hat{N} := \frac{Q}{g} ~~,~~
\hat{x}:= x g~~,~~
\hat{t} := t g~~,~~
\hat{L} := L g ~~,~~
\hat{m} := \frac{e^{\gamma_E}}{\pi g}m_\psi~~,
\end{align}
so that the equations of motion become
\begin{align}
(\hat{\partial}^2 + 1)\vf_c &= \hat{N} (\theta(\hat{x}-\hat{L}/2) - \theta(\hat{x}+\hat{L}/2) 
+ \hat{m} \sin(2 \sqrt{\pi} \vf_c + \theta).
\label{eqMD}
\end{align}
Fig.~\ref{Fig2} depicts final fields profile for $\vf_c$ that solve the above equation of motion at $\theta = 0$ and $\hat{N} = 200$, for a range of values of $\hat{L}$ and $\hat{m}$. 

Before discussing the obtained solutions we briefly summarize our numerical algorithm.\footnote{Our codes are available at \url{https://github.com/valerivardanyan/QED-discharge}.} The final field profile is a solution of an elliptical boundary value problem. The equation of motion in our case is non-linear for non-zero values of $\hat{m}$, and we are using an iterative relaxation method to find its solution. We start with a proposal profile, linearize the equation of motion in terms of deviation from the proposal profile, and solve for linear deviations. We then iterate this procedure until the updated profile solves the non-linear equation with sufficient precision. Such an algorithm is commonly used for handling highly non-linear boundary value problems. More details can be found in e.g. \cite{symmetron_numerics}. The results of the numerical integration is plotted in Fig. \ref{Fig2}, and depicts the fully discharged electric-field for various values of $\hat{m}$, and taking $\hat{N} = 200$ for concreteness. We see that the final discharged E-field profile for massless QED\t, and those for massive QED\t, are very close to each-other for $\hat{m} \ll \hat{N}$.

As $\hat{m}/\hat{N}$ increases, the source and the potential terms in the classical equation of motion start to become comparable to each other, culminating in fact that when $\hat{m}/\hat{N} = 2$, the maximally discharged state of the capacitor is very close to the initial state of the system, as can be seen from Fig. \ref{Fig2}. This is evidently the (1+1)-$d$ manifestation of the mass threshold seen in (3+1)-$d$, which exponentially cuts off as the mass of the charged quanta increases according to Eq. \eqref{pp}: $\Gamma \propto e^{-\frac{\pi m^2}{e E}}$. Here, the approach to the threshold evidently proceeds linearly as $\hat{m}/\hat{N} \to 2$ and cuts off at $\hat m/\hat{N} = e^{\gamma_E} m_\psi/(\sqrt\pi Q) = 2$, implying a critical electric field of 
\eq{mthresh}{E_c = \frac{e^{\gamma_E}m_\psi}{2\sqrt\pi} \approx \frac{m_\psi}{2}}
below which discharge effectively does not occur.\footnote{Note that in (1+1)-$d$, electric fields and charge both have mass dimension one, to be contrasted to (3+1)-$d$ where $E$ has dimension mass squared and charge is dimensionless. The mass threshold in Eq. \eqref{mthresh} is only valid within our approximation of large occupation number, i.e. when the initial electric field $E_0 \gg e$. For much smaller values of $E_0$ beyond the remit of our approximation, one can still show that discharge does not occur unless $E_0 > \frac{e}{2}$ \cite{Coleman:1976uz}.} It would be a mistake, however, to interpret the bare $m_\psi$ as the mass of any physical states of the massive Schwinger model. To understand why, we need to understand the spectrum of states in the theory.

The bosonized massive Schwinger model is qualitatively distinct from the sine-Gordon model in that there are no degenerate vacua for non-vanishing electric charge with the choice $\theta = 0$. This is due to the presence of the `mass' term in Eq. \eqref{bseom} proportional to $g^2$ which gaps the spectrum even for the bosonized version of the massless Schwinger model. This makes sense, as there are no free charges in the theory, and the kink/ anti-kink solutions in the sine-Gordon model are the solitonic dual of free fermions. Since the Schwinger model can only have bound states of these fermions in its spectrum, their bosonic dual states must be more complicated. Recalling the result for the spectrum of stable states of the massive Schwinger model in the regime where $m_\psi/g  = \hat{m} \gg 1$~\cite{Coleman:1976uz}, we see that when $\theta$ is set equal to zero, there are in fact $N_s \approx \hat{m}^2$ stable bound-states with masses $m \leq 4 m_\psi$ that scale as
\begin{align}
m_n \approx \Delta \sqrt{n}  ~,~{\rm for}~\Delta = 4 g~,~{\rm and} ~1 \leq n \leq N_s \approx \hat{m}^2 = \tfrac{m_\psi^2}{g^2} \implies 
\begin{cases}
m_1 ~~= m_{\rm min} &\!\!\!\approx 4 g ~,\\
m_{N_s} = m_{\rm max} &\!\!\! \approx 4 m_\psi~.
\end{cases}
\end{align}
Therefore the $\p$ field in the bosonized theory must correspond to a collective excitation of this finite tower of meson states, with lightest mass $4g$ and heaviest mass $4m_\psi$ and as such, cannot cleanly be identified with a single mass scale.

Now, recall the dimensional analysis in section~\ref{secClouds}. This analysis tied the exponential nature of the small-tails in the fully discharged E-field in massless QED\t~ rigidly to the Compton wavelength of the massive $\vf$-boson, $\ell_\vf = 1/g$; this is well-motivated, as the $\vf$-meson is the only on-shell degree of freedom in massless QED\t. Naively, were we to apply this same logic to massive QED\t, we would expect that shape of the final-state E-fields for large $\hat{m}$ is similarly dictated by the Compton wavelengths of these stable particles. In particular, because the lightest stable meson has a mass $m \approx 4g$, one might expect roughly similar field profiles. 

However, as is clearly seen in Fig.~\ref{Fig2}, this analysis does not describe the shape of the fully discharged E-fields for massive QED\t~whenever $\hat{m} = m_\psi/g \gg 1$, particularly for the plate separation threshold evident on the leftmost panel. One possible interpretation for this, is that pair producing mesons from the vacuum still requires overcoming the mass thresholds for the constituent fermions, before flux tubes form and compensate the bound state energy, resulting in the composite state having a mass significantly less than the sum of its constituents. In this way, one still has to overcome a large mass threshold to produce these bound states, even if the resulting condensate consists of parametrically lighter states. The dipole length commensurate with the lightest of these states with mass $m_1$ may be parametrically quite different from $m^{-1}_1$ given the large binding energies involved for larger values of the bare mass, resulting in $\ell_\vf \ll 1/m_1$ for larger values of the bare mass and approaching $\ell_\vf \sim 1/m_1 \sim 1/g$ for smaller values of the bare mass, which would simultaneously account for the mass and plate separation thresholds depicted in Fig. \ref{Fig2}. However, a quantitative understanding of this is beyond the scope of this paper and necessitates further investigation. Here, we merely wish to point out the existence of the independent mass and plate separation thresholds in massive QED\t, a conclusion we can confidently draw from the accuracy of integrating the classical equations of motion as an approximation of the bosonized description at large occupation number.

Finally, it is also interesting to note that for fermion masses in the lighter range ($m_\psi/g \lesssim \calO(1)$), the consistency of our approximation results in the source term dominating the dynamics, and so the discharge proceeds almost as in the massless case, in which care we can consider the analogs of the timescale in the massless QED\t, but now with the degeneracy between $g$ and $m_\vf$ broken according to Eq. \eqref{mm}, so that the three timescales associated with discharge from Eqs.~\eqref{eqTZ} and~\eqref{eqTVV} become
\begin{align}
\tz &= \frac{1}{g},\\
\tvv &=\frac{1}{m_\vf} ~\frac{Q^2 L^2}{\pi} \left(\frac{g}{m_\vf}\right)^6 \left(1 - \frac{1-e^{-m_\vf L}}{m_\vf L}\right)^2, \\
\tvs &= \frac{1}{g}~\frac{g^2 L^2}{4 \pi} \left(\frac{g}{m_\vf}\right)^3 ,
\end{align}
allowing for parametrically different behavior for the late vs. early time dynamics of the discharge relative to the massless case. A complete understanding of the dynamics of capacitor discharge (in particular with small initial charge) is not possible within our approximation.\footnote{A notable attempt to study electric field discharge in the massive Schwinger model was made using lattice simulations in \cite{Hebenstreit:2013qxa} building on the formalism introduced in \cite{Aarts:1998td}, however the computational cost of these simulations meant that only the initial stages of this discharge could be tracked.}

\section{Discharge in higher dimensions ($d > 2$)}\label{secGen}

When viewed as a tunneling process, the salient aspects of Schwinger pair production in (3+1)-dimensions are often well-captured by computing the probability of filled, negative energy states tunneling through an effective potential-barrier in (1+1)-dimensions, as formulated in e.g.~\cite{Nikishov:1970br,Cohen:2008wz,Allor:2007ei}. It is thus tempting to use our fully back-reacted discharge in (1+1)-dimensions to model capacitor-discharge in higher-dimensions. However, there are prominent obstacles between using our present understanding of capacitor discharge in massless QED$_2$ and any full understanding of capacitor discharge in higher-dimensions, whose root cause come from the computation of the tunneling process to begin with.

The main, essential, difficulty here is the notion of scattering states themselves. The simpler case of electron-positron pairs creation due to a \textit{static} electric field between two charged capacitor plates can easily be reframed in terms of an integral over a continuum of scattering states which tunnel through an effective potential. Explicitly, as the capacitor plates are invariant in the $d-2$ transverse directions, then the problem becomes (1+1)-dimensional. Further, because the E-field is static in time, then by choosing the gauge where $\vec{E} = - \vec{\nabla} \phi$, the equations of motion are time-translation invariant. Thus, here, one can define scattering states with well-defined energy $E$ and transverse momentum $\vec{p}_T$, incident on the effective potential set-up by the E-field between the capacitor plates~\cite{Nikishov:1970br,Wang:1988ct}. It is then straightforward to find the tunneling amplitudes and thus the rate of pair production as discussed in e.g.~\cite{Cohen:2008wz,Allor:2007ei,McGady:2017slr}. 

Unfortunately, the very fact that the E-field is static at once makes this set-up tractable while also excising any appreciable back-reaction. (Formally, this setup describes Schwinger pair-creation in the limit where the electron charge is sent to zero whilst the field-strength is held fixed, i.e. where $e \to 0$ while $e E$-fixed.) Relaxing this condition and allowing the E-field to vary in time, i.e. to allow it to \textit{discharge}, ruins the time-translation invariance needed to label the scattering states by their energy and, further, causes the ``effective potential'' to vary in time. Together, these two technical problems obfuscate the most obvious paths between our exact treatment for (1+1)-dimensional discharge and understanding discharge in higher dimensions.

Related to this, in $d > 2$ there are nontrivial magnetic fields. Thus, the time-variation of the electric field induces magnetic fields. These magnetic fields deflect the paths of charged particles and paths between our exact solution in (1+1)-dimensions and exact understandings of capacitor discharge in $d$-dimensions alike. For these reasons, we expect the results/techniques in this paper may be of limited use in understanding detailed dynamics of capacitor discharge above $d = 2$. 

\section{Discussion}\label{secEnd}

Having reinterpreted and generalized the findings of \cite{Chu:2010xc} to incorporate bare fermion masses within the Schwinger model, the main findings of our investigation can be summarized as:
\begin{itemize}
	\item Electric fields can discharge even in theories exhibiting confinement via dipole production. These correspond to meson bound states in the context of the Schwinger model.
	\item This dipole production will necessarily be suppressed below the threshold plate separation corresponding to the characteristic size of the dipole $L = \ell_\p \sim g^{-1}$ in the massless case.
	\item When generalized to the massive case, the mesons correspond to a finite tower bound states, and the threshold separation becomes a more complicated function of the bare fermion mass and $g$. 
	\item The introduction of bare fermion masses to the Schwinger model results in the appearance of an additional threshold at $E_c = \frac{e^{\gamma_E}m_\psi}{2\sqrt\pi} \approx \frac{m_\psi}{2}$, below which electric field discharge is suppressed.	
\end{itemize}
There are a number of conceptual issues, however, that remain to be discussed. The issue of how to realize the initial conditions of two charges of $\pm Q$ separated by a distance $L$ \textit{within} massless QED\t~--- i.e. how to construct the capacitor itself directly in terms of polar $\vf$-mesons themselves ---  is a very interesting and, to our knowledge, open question. It should be possible to directly construct this initial condition as coming from a specific initial state of free, non-interacting, $\vf$-mesons as the bosonized theory is free and integrable. Furthermore, it would be useful to understand whether quantitative progress can be made away from our semi-classical approximation which relied on large initial occupation, and therefore large initial electric fields. However carrying out these exercises is beyond the scope of the current work. Here, we content ourselves with the observation that at large occupation number, provided the initial capacitor state can be constructed (possibly involving operators in a larger theory, a consistent truncation of which is the Schwinger model), charge conservation and the unitarity of the Schwinger model renders our treatment self-consistent.

\section*{Acknowledgements}
We'd like to thank Konstantin Zarembo, Tanmay Vachaspati, and Poul Henrik Daamgard for insightful and helpful comments on the draft. DM was supported by the grant ``Exact Results in Gauge and String Theories'' from the Knut and Alice Wallenberg foundation.  V.V.  is supported by the WPI Research Center Initiative,  MEXT,  Japan,  and by JSPS KAKENHI Grant Number 20K22348.   
\appendix

\end{document}